\documentclass[prb,reprint,showpacs,twocolumn,superscriptaddress,floatfix]{revtex4}
\usepackage{epsfig}
\usepackage{amsmath}
\usepackage{amssymb}
\usepackage{amsfonts}
\usepackage{mathptmx}
\usepackage{dcolumn}
\usepackage{eucal}
\usepackage{bm}
\usepackage{color}
\usepackage[colorlinks,linkcolor=blue,citecolor=blue]{hyperref}

\usepackage{epstopdf}

\usepackage{graphicx}

\usepackage{natbib}

\begin{document}
\title{Coherent diffraction of thermal currents in Josephson tunnel junctions}
\author{F. Giazotto}
\email{giazotto@sns.it}
\affiliation{NEST, Instituto Nanoscienze-CNR and Scuola Normale Superiore, I-56127 Pisa, Italy}
\author{M. J. Mart\'inez-P\'erez}
\affiliation{NEST, Instituto Nanoscienze-CNR and Scuola Normale Superiore, I-56127 Pisa, Italy}
\author{P. Solinas}
\affiliation{Laboratoire Mat\'eriaux et Ph\'enom\`enes Quantiques, Universit\'e Paris Diderot-Paris 7 and CNRS, UMR 7162, 75013 Paris, France}
\affiliation{SPIN-CNR, Via Dodecaneso 33, 16146 Genova, Italy}

\pacs{74.50.+r,74.25.F-,74.25.fc,85.80.Fi}
\begin{abstract}
We theoretically investigate heat transport in temperature-biased Josephson tunnel junctions in the presence of an in-plane magnetic field. 
In full analogy with the Josephson  critical current,
the phase-dependent component of the heat flux through the junction displays \emph{coherent diffraction}. 
Thermal transport is analyzed in three prototypical junction geometries highlighting their main differences. Notably,
minimization of the Josephson coupling energy 
requires the quantum phase difference across the junction to undergo $\pi$ \emph{slips} in suitable intervals of magnetic flux. 
An experimental setup suited to detect thermal diffraction is proposed and analyzed.    
\end{abstract}

\maketitle
\section{Introduction}
\label{intro}
The impressive advances achieved in nanoscience and technology are nowadays enabling the understanding of one  central topic in science, i.e., \emph{thermal flow}   in solid-state nanostructures \cite{Giazotto2006,Dubi2011}.  
Control and manipulation \cite{heattransistor,ser} of thermal currents in combination with the investigation of the origin of dissipative phenomena are of particular relevance at such scale where heat deeply affects the properties of the systems, for instance,
from \emph{coherent caloritronic} circuits, which allow enhanced operation thanks to the quantum phase \cite{Meschke2006,Vinokur2003,Eom1998,Chandrasekhar2009,Ryazanov1982,Panaitov1984,virtanen2007,Martinez2013}, to more developed research fields such as ultrasensitive radiation detectors \cite{Giazotto2006,Giazotto2008} or cooling applications \cite{Giazotto2006,Giazotto2002}.
In this context it has been known for more than $40$ years that heat transport in Josephson junctions can be, in principle, phase-dependent \cite{Maki1965,Guttman97,Guttman98,Zhao2003,Zhao2004,Golubev2013}.   
The first ever Josephson thermal interferometer has been, however, demonstrated only very recently \cite{giazotto2012,martinez2012,giazottoexp2012,simmonds2012},
therefore proving that phase coherence extends to thermal currents as well.   
The heat interferometer of Ref. \cite{giazottoexp2012}  might represent a  prototypical circuit to implement  novel-concept coherent caloritronic devices such as heat transistors \cite{martinez2012}, thermal splitters and rectifiers \cite{Martinez2013}.   

In the present work we theoretically analyze heat transport in temperature-biased extended Josephson tunnel junctions showing that the phase-dependent component of thermal flux  through the weak-link interferes in the presence of an in-plane magnetic field  leading to \emph{heat diffraction}, in analogy to what occurs for the Josephson critical current.
In particular, thermal transport is investigated in three prototypical \emph{electrically-open} junctions geometries showing that the quantum phase difference across the junction undergoes $\pi$ \emph{slips} in order to minimize the Josephson coupling energy.
These phase slips have energetic origin and are not related to fluctuations as conventional phase slips in low-dimensional superconducting systems \cite{Langer1967, Zaikin1997,astafiev2012}.
We finally  propose how to demonstrate thermal diffraction in a realistic microstructure,  and to prove such $\pi$ slips exploiting an uncommon observable such as the heat current.

The paper is organized as follows: In Sec. \ref{model} we describe the general model used to derive the behavior of the heat current in a temperature-biased extended Josephson tunnel junction. 
In Sec. \ref{results} we obtain the conditions for the quantum phase difference across an electrically-open short Josephson junction in the presence of an in-plane magnetic field, and the resulting behavior of
 the phase-dependent thermal current. In particular, we shall demonstrate the occurrence of phase-slips of $\pi$, independently of the junction geometry, in order to minimize the Josephson coupling energy. The phase-dependent heat current in three specific junction geometries is further analyzed in Sec. \ref{differentgeo}, where we highlight their main differences. 
In Sec. \ref{experiment} we suggest and analyze a possible experimental setup suited to detect heat diffraction through electronic temperature measurements in a microstructure based on an extended Josephson junction, and to demonstrate the existence of $\pi$ slips.
Finally, our results are summarized in Sec. \ref{summary}.

\begin{figure}[t!]
\includegraphics[width=\columnwidth]{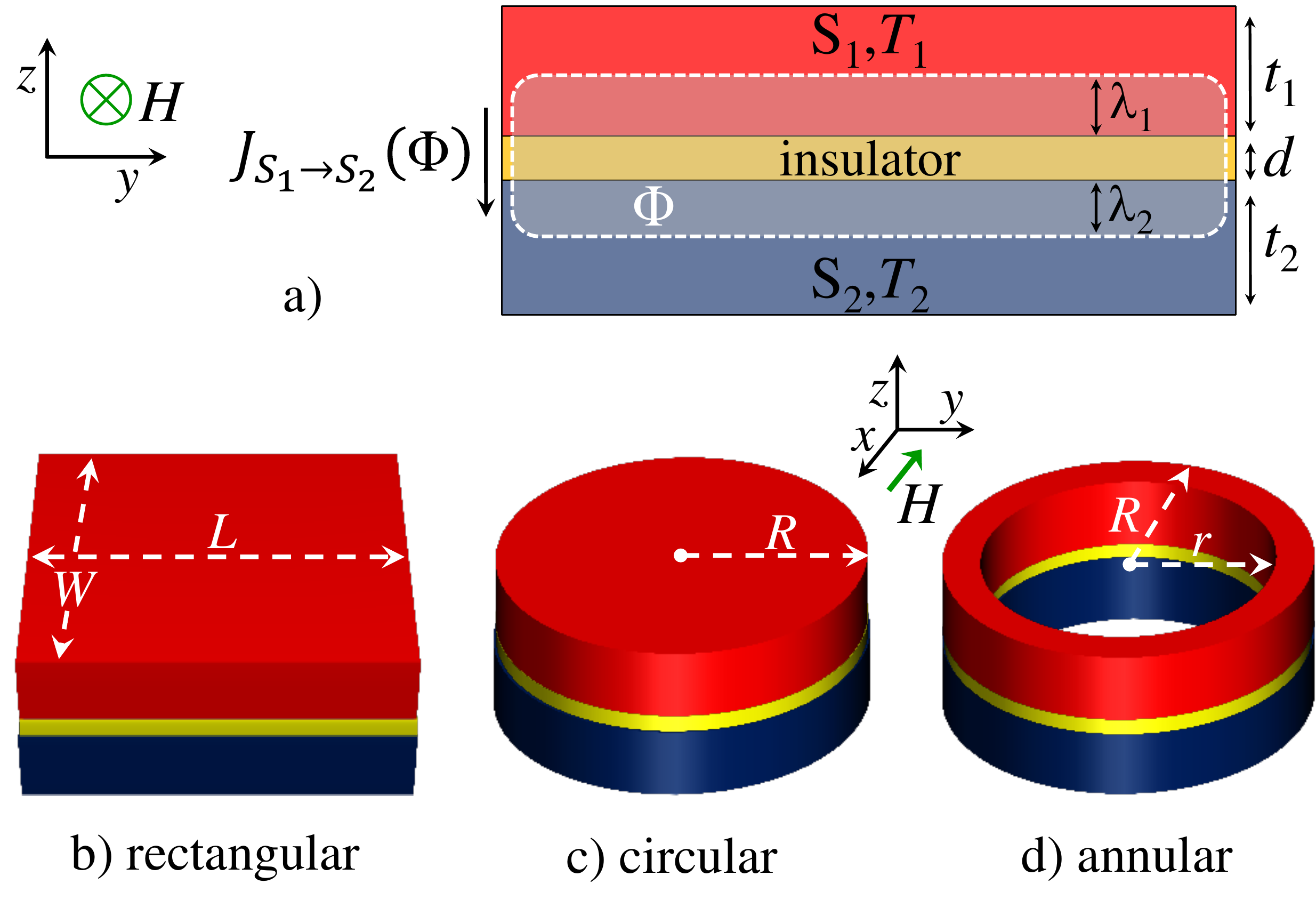}
\caption{(Color online) (a) Cross section of a temperature-biased extended  S$_1$IS$_2$ Josephson tunnel junction in the presence of an in-plane magnetic field $H$. 
The heat current $J_{S_1\rightarrow S_2}$ flows along the $z$ direction whereas $H$ is applied in the $x$ direction, i.e., parallel to a symmetry axis of the junction. 
Dashed line indicates the closed integration contour, $T_i$, $t_i$ and $\lambda_i$ represent  the temperature, thickness and London penetration depth of superconductor S$_i$, respectively, and $d$ is the insulator thickness. $\Phi$ denotes the magnetic flux piercing the junction.  
Prototypical junctions with rectangular, circular, and annular geometry are shown in panel (b), (c) and (d), respectively. $L$, $W$, $R$ and $r$ represent the junctions geometrical parameters. 
}
\label{fig1}
\end{figure}
\section{Model}
\label{model}
Our system is schematized in Fig. \ref{fig1}(a), and consists of an extended Josephson tunnel junction composed of two superconducting electrodes S$_1$ and S$_2$ in thermal steady-state residing at different temperatures $T_1$ and $T_2$, respectively. 
We shall focus mainly on symmetric Josephson junctions in the \emph{short} limit, i.e., with lateral dimensions much smaller than the Josephson penetration depth [see Fig. \ref{fig1}(b,c,d)], $L,W,R,r\ll \lambda_J=\sqrt{\frac{\pi \Phi_0}{\mu_0i_ct_H}}$, where $\Phi_0=2.067\times 10^{-15}$ Wb is the flux quantum, $\mu_0$ is vacuum permeability, $i_c$ is the critical current areal density of the junction, and $t_H$ is the junction effective magnetic thickness to be defined below.
In such a case the self-field generated by the Josephson current in the weak-link can be neglected with respect to the externally applied magnetic field, and no traveling solitons can be originated. 
$t_i$ and $\lambda_i$ denote the thickness and London penetration depth of superconductor S$_i$, respectively, whereas $d$ labels the insulator thickness. We choose a coordinate system such that the applied magnetic field ($H$) lies parallel to a symmetry axis of the junction and along $x$, and that the junction electrodes planes are parallel to the $xy$ plane. 
Furthermore, the junction lateral dimensions are assumed to be much larger than $d$ so that we can neglect the effects of the edges, and
each superconducting layer is assumed to be thicker than its London penetration  depth (i.e., $t_i>\lambda_i$) so that $H$ will penetrate the junction in the $z$ direction within a thickness $t_H=\lambda_1+\lambda_2+d$ \cite{magneticlength}.
For definiteness, we assume $T_1\geq T_2$ so that the Josephson junction is temperature biased only,  and no electric current flows through it. If $T_1\neq T_2$ there is a finite electronic heat current $J_{S_1\rightarrow S_2}$ flowing through the junction from S$_1$ to S$_2$ [see Fig. \ref{fig1}(a)] which is given by \cite{Maki1965,Guttman97,Guttman98,Zhao2003,Zhao2004,Golubev2013}
\begin{equation}
J_{S_1\rightarrow S_2}(T_1,T_2,\varphi)=J_{qp}(T_1,T_2)-J_{int}(T_1,T_2)\textrm{cos}\varphi.
\label{heatcurrent}
\end{equation} 
Equation (\ref{heatcurrent}) describes the oscillatory behavior of the thermal current flowing through a Josephson tunnel junction as a function of $\varphi$ predicted by Maki and Griffin \cite{Maki1965}, and experimentally verified in Ref. \cite{giazottoexp2012}.
In Eq. (\ref{heatcurrent}),  $J_{qp}$ is the usual heat flux carried by quasiparticles \cite{Giazotto2006,Frank1997}, $J_{int}$ is the phase-dependent part of the heat current which is peculiar of Josephson tunnel junctions, and $\varphi$ is the macroscopic quantum phase difference between the superconductors. By contrast, the Cooper pair condensate does not contribute to heat transport in a static situation \cite{Maki1965,Golubev2013,giazottoexp2012}.
 
The two terms appearing in Eq. (\ref{heatcurrent}) read \cite{Maki1965,Guttman97,Guttman98,Zhao2003,Zhao2004,Golubev2013}
\begin{equation}
J_{qp}=\frac{1}{e^2 R_J} \int^{\infty}_{0} d\varepsilon \varepsilon \mathcal{N}_1 (\varepsilon,T_1)\mathcal{N}_2 (\varepsilon,T_2)[f(\varepsilon,T_2)-f(\varepsilon,T_1)], 
\end{equation}
and
\begin{equation}
J_{int}=\frac{1}{e^2 R_J}\int^{\infty}_{0}d\varepsilon \varepsilon \mathcal{M}_{1}(\varepsilon,T_1)\mathcal{M}_{2}(\varepsilon,T_2)[f(\varepsilon,T_2)-f(\varepsilon,T_1)],
\end{equation}
where $\mathcal{N}_{i}(\varepsilon,T_i)=|\varepsilon|/\sqrt{\varepsilon^2-\Delta_{i}(T_i)^2}\Theta[\varepsilon^2-\Delta_{i}(T_i)^2]$ 
is the BCS normalized  density of states in S$_{_i}$ at temperature $T_i$ ($i=1,2$),
$\mathcal{M}_{i}(\varepsilon,T_i)=\Delta_{i}(T_i)/\sqrt{\varepsilon^2-\Delta_{i}(T_i)^2}\Theta[\varepsilon^2-\Delta_{i}(T_i)^2]$,
and $\varepsilon$ is the energy measured from the condensate chemical potential.
Furthermore,
$\Delta_i(T_i)$ is the temperature-dependent superconducting energy gap, $f(\varepsilon,T_i)=\text{tanh}(\varepsilon/2 k_BT_i)$,
$\Theta(x)$ is the Heaviside step function,
$k_B$ is the Boltzmann constant, $R_J$ is the junction normal-state resistance,
and $e$ is the electron charge.
In the following analysis we neglect any contribution to thermal transport through the Josephson junction arising from lattice phonons.

\section{Results}
\label{results}

In order to discuss the effect of the applied magnetic field on the heat current we shall focus first of all onto the phase-dependent component. 
To this end we need to determine the phase gradient $\varphi (x,y)$ induced by the application of the external magnetic flux. 
By choosing the closed integration contour indicated by the dashed line depicted in Fig. \ref{fig1}(a)
it can be shown \cite{Tinkham,Barone} that, neglecting screening induced by the Josephson current, $\varphi (x,y)$ obeys the equations $\partial \varphi/\partial x=0$ and $\partial \varphi/\partial y=2\pi \mu_0 t_H H/\Phi_0$. The latter equation can be easily integrated to yield  
\begin{equation}
\varphi (y)=\kappa y+\varphi_0, 
\end{equation}
where $\kappa \equiv 2\pi \mu_0 t_H H/\Phi_0$ and $\varphi_0$ is the phase difference at $y=0$. 
The phase-dependent component of the heat current can  then be written as
\begin{equation}
J_{H}(T_1,T_2,H)=\int\int dxdy J_A(x,y,T_1,T_2)\textrm{cos}(\kappa y+\varphi_0),
\label{phasea}
\end{equation}
where the integration is performed over the junction area, and $J_A(x,y,T_1,T_2)$ is the heat current density per unit area. 
We note that the integrand of Eq. (\ref{phasea}) oscillates sinusoidally along the $y$ direction with period given by $\Phi_0(\mu_0t_H H)^{-1}$.
After integration over $x$ we can write Eq. (\ref{phasea}) as 
\begin{eqnarray}
J_{H}(T_1,T_2,H)&=&\int dy \mathcal{J}(y,T_1,T_2)\textrm{cos}(\kappa y+\varphi_0)\nonumber\\
&=&\textrm{Re}\left\{e^{i\varphi_0}\int^{\infty}_{-\infty} dy \mathcal{J}(y,T_1,T_2)e^{i\kappa y}\right\},
\label{phaseb}
\end{eqnarray}  
where $\mathcal{J}(y,T_1,T_2)\equiv \int dx J_A(x,y,T_1,T_2)$ is the heat current density per unit length along $y$. 
In writing  second equality in Eq. (\ref{phaseb}) we have replaced the integration limits by $\pm \infty$ since the thermal current is zero outside the junction.
Equation (\ref{phaseb}) for $J_{H}(T_1,T_2,H)$ resembles the expression for  the Josephson current, $I_{H}(T_1,T_2,H)$, which is given by \cite{Tinkham,Barone}
\begin{eqnarray} 
I_{H}(T_1,T_2,H)&=&\textrm{Im}\left\{e^{i\varphi_0}\int^{\infty}_{-\infty} dy \mathcal{I}(y,T_1,T_2)e^{i\kappa y}\right\} \nonumber\\
&=&\textrm{sin}\varphi_0\int^{\infty}_{-\infty} dy\mathcal{I}(y)\textrm{cos}\kappa y,
\label{critical}
\end{eqnarray}
where $\mathcal{I}(y,T_1,T_2)$ is the supercurrent density integrated along $x$,
and second equality in Eq. (\ref{critical}) follows from the assumed junctions \emph{symmetry}, i.e., $\mathcal{I}(y,T_1,T_2)=\mathcal{I}(-y,T_1,T_2)$.
In the actual configuration of electrically-open junction, the condition of \emph{zero} Josephson current for any given value of $H$
yields the solution $\varphi_0=m\pi$, with $m=0,\pm 1,\pm2\ldots$.  
On the other hand, the Josephson coupling energy of the junction ($E_J$) can be expressed as
\begin{eqnarray}
E_J(T_1,T_2,H)&=&E_{J,0}-\frac{\Phi_0}{2\pi}\textrm{Re}\left\{e^{i\varphi_0}\int_{-\infty}^{\infty} dy \mathcal{I}(y,T_1,T_2)e^{i\kappa y}\right\}\nonumber\\ 
&=&E_{J,0}-\frac{\Phi_0}{2\pi}\textrm{cos}\varphi_0\int_{-\infty}^{\infty} dy\mathcal{I}(y)\textrm{cos}\kappa y
\label{energy}
\end{eqnarray}
where
$E_{J,0}=\Phi_0I_c/2\pi$, $I_c$ is the zero-field critical supercurrent,
and in writing the second equality we have used the symmetry property of $\mathcal{I}(y,T_1,T_2)$. 
Minimization of $E_J$ for any applied $H$ imposes the second term on rhs of Eq. (\ref{energy}) to be always negative, so that
 $\varphi_0$ will undergo a $\pi$ \emph{slip} whenever the integral does contribute to $E_J$ with negative sign. 
As a result, the Josephson  coupling energy turns out to be written as
\begin{equation}
E_J(T_1,T_2,H)=E_{J,0}-\frac{\Phi_0}{2\pi}\left|\int_{-\infty}^{\infty} dy\mathcal{I}(y,T_1,T_2)\textrm{cos}\kappa y\right|. 
\label{energybis}
\end{equation} 
We also assume that the symmetry of the junction and of the electric current  density  are reflected in an analogous symmetry in the heat current, i.e., $\mathcal{J}(y,T_1,T_2) = \mathcal{J}(-y,T_1,T_2)$.
It therefore follows from Eq. (\ref{phaseb}) that $J_H$ can be written as
\begin{equation}
J_{H}(T_1,T_2,H)=\left|\int^{\infty}_{-\infty} dy \mathcal{J}(y,T_1,T_2)\textrm{cos}\kappa y\right|.
\label{heatcurrfin}
\end{equation}
Equation (\ref{heatcurrfin}) is the main result of the paper.

The above results hold for symmetric Josephson junctions under in-plane magnetic field parallel to a symmetry axis, and only occur without any electrical bias. 
The discussed phase slips, however, exists for any arbitrary junction geometry.
If the junction has not a symmetric geometry with respect to the magnetic field direction, the constraints on vanishing Josephson current and minimization of coupling energy are translated in a more complex condition for the phase $\varphi_0$.
As we shall demonstrate below, the phase undergoes nevertheless a $\pi$ slip as well.
We choose the $x$ axis of the coordinate system parallel to the magnetic field [as in Fig. \ref{fig1}(a)].
For a junction with arbitrary symmetry, we split $\mathcal{I}(y,T_1,T_2)$ in its symmetric, $ \mathcal{I}_s(y)$, and antisymmetric part, $ \mathcal{I}_a(y)$. We thus have
\begin{eqnarray}
 I_H&=&\textrm{Im}\left\{e^{i\varphi_0}\int^{\infty}_{-\infty} dy \mathcal{I}(y)e^{i\kappa y}\right\} \nonumber\\
&=&  \cos \varphi_0  \int_{-\infty}^{\infty} dy \mathcal{I}_a (y) \sin  \kappa y + \sin \varphi_0  \int_{-\infty}^{\infty} dy \mathcal{I}_s (y) \cos  \kappa y\nonumber\\
 &=& \cos \varphi_0 I_a + \sin \varphi_0 I_s,
 \label{eq:zero_current}
\end{eqnarray}
where we have denoted the symmetric and antisymmetric integrals as $I_s$ and $I_a$, respectively.
The case of symmetric junctions has already been discussed above so that in the following we shall focus on junctions with no symmetry, i.e., with $I_s\neq0$  {\it and} $I_a\neq0$. Notice that this already has consequences on the values of $\varphi_0$.
In fact, if $I_s\neq0$  {\it and} $I_a\neq0$ we must have $\cos \varphi_0 \neq 0$ {\it and} $\sin \varphi_0 \neq 0$ to satisfy the zero-current condition.

The Josephson coupling energy can be written as 
\begin{eqnarray}
E_J&=&E_{J,0} - \frac{\Phi_0}{2 \pi}  \Big [\cos \varphi_0  \int_{-\infty}^{\infty} dy \mathcal{I}_s (y) \cos  \kappa y\nonumber\\
&-& \sin \varphi_0  \int_{-\infty}^{\infty} dy \mathcal{I}_a (y) \sin  \kappa y \Big ]\nonumber\\ 
&=&E_{J,0} + \frac{\Phi_0}{2 \pi}  \Big [ -\cos \varphi_0 I_s + \sin \varphi_0 I_a \Big].
\end{eqnarray}
To find the energy minima, we differentiate twice with respect to $\varphi_0$ and impose the condition  $\partial^2 E_J / \partial \varphi_0^2 > 0$. We therefore obtain
\begin{equation}
  \frac{\partial^2 E_J}{ \partial \varphi_0^2} =  \frac{\Phi_0}{2 \pi}  \Big [ \cos \varphi_0 I_s - \sin \varphi_0 I_a \Big] >0.
 \label{eq:energy_min}
\end{equation}
Assuming that that $I_s \neq 0$, the condition of vanishing $I_H$ for any applied $H$ from Eq. (\ref{eq:zero_current}) reads
\begin{equation}
 \sin \varphi_0 = - \cos \varphi_0 \frac{I_a}{I_s} ~ {\rm or }~ \tan \varphi_0 = - \frac{I_a}{I_s}.
\label{eq:sin_varphi_0}
\end{equation}
By using the first of Eqs. (\ref{eq:sin_varphi_0}), the condition to have minima in Eq. (\ref{eq:energy_min}) gives
\begin{equation}
 \frac{\partial^2 E_J}{ \partial \varphi_0^2} = \frac{\Phi_0}{2 \pi}  \cos \varphi_0 \Big(I_s + \frac{I_a^2}{I_s} \Big) = \frac{\Phi_0}{2 \pi}  \frac{\cos \varphi_0}{I_s} \left(I_s^2 + I_a^2 \right)  > 0
\label{eq:min_condition}
\end{equation}
that depends on the signs of $I_s$ and $\cos \varphi_0$.

Now we turn to equations (\ref{eq:sin_varphi_0}) that impose a constraint on $\varphi_0$ as a function of $I_s$ and $I_a$.
To simplify the discussion we denote as $\phi_0 = - \arctan (I_a/I_s)$, and consider only solutions within a $2\pi$ variation from the latter.
We have two solutions of Eqs. (\ref{eq:sin_varphi_0}): $\varphi_{0,1} = \pi + \phi_0$ and $\varphi_{0,2} =  \phi_0$ which correspond to cosine function
\begin{eqnarray}
 \cos \varphi_{0,1} &=& - \cos \phi_0 =  - \frac{1}{\sqrt{1+ \left(\frac{I_a}{I_s} \right)^2}} \nonumber \\
  \cos \varphi_{0,2} &=&  \cos \phi_0 =  \frac{1}{\sqrt{1+ \left(\frac{I_a}{I_s} \right)^2}},
\end{eqnarray}
where we have used the relation $\cos( \arctan x)= 1/\sqrt{1+x^2}$.
As we can see, the first solution gives a negative $\cos \varphi_{0,1}$ while the second one corresponds to positive $\cos \varphi_{0,2}$.
Going back to the inequality (\ref{eq:min_condition}), if $I_s>0$ we need to choose the solution $\varphi_{0,2} =  \phi_0$ (for which $ \cos \varphi_{0,2}>0$)
 to minimize the Josephson coupling energy.
By contrast, if $I_s<0$, we must choose the solution $\varphi_{0,1} = \pi + \phi_0$ (for which $ \cos \varphi_{0,1}<0$).
Therefore, we get that the superconducting phase must undergo a $\pi$ slip to minimize the Josephson coupling energy whenever the integral $I_s$ changes sign as a function of the magnetic field.

We shall conclude by discussing the pure \emph{antisymmetric} junction case, i.e., $I_a\neq0$ and $I_s =0$.
Because of  the zero current condition the only values that the phase $\varphi_0$ can assume are $\pi/2$ or  $3\pi/2 $.
 Equation (\ref{eq:energy_min}) implies that, if $I_a >0$, $\varphi_0=3\pi /2$ and, if $I_a <0$, $\varphi_0=\pi/2$. Therefore, also in this case, the phase $\varphi_0$ undergoes a $\pi$ slip when $I_a$ changes sign.



We remark that the discussed phase slips differ from those present in low dimensional superconductors,
caused by thermal \cite{Langer1967} and quantum \cite{Zaikin1997,astafiev2012} fluctuations.
In those cases, the phase slips are generated when, because of fluctuations, the modulus of the complex order parameter goes to zero, the phase becomes unrestricted and jumps of $2 \pi$ \cite{arutyunov08}.

By contrast, the phase slips discussed above have an energetic origin and they occur when the system passes from one energetically stable configuration to another one \cite{kuplevakhsky06}.
This transition takes place when the magnetic flux crosses one of the critical points and therefore can be experimentally induced by changing the magnetic flux.
The different origin of the slips is exemplified by the fact that the fluctuation-induced phase slips are always of $2 \pi$ while in the present case we have slips of $\pi$.
The identification of this effect is possible only in the electrically-open junctions. In fact, the presence of an electric current or a voltage bias would destroy or hide the  original effect.

\begin{figure}[t!]
\includegraphics[width=\columnwidth]{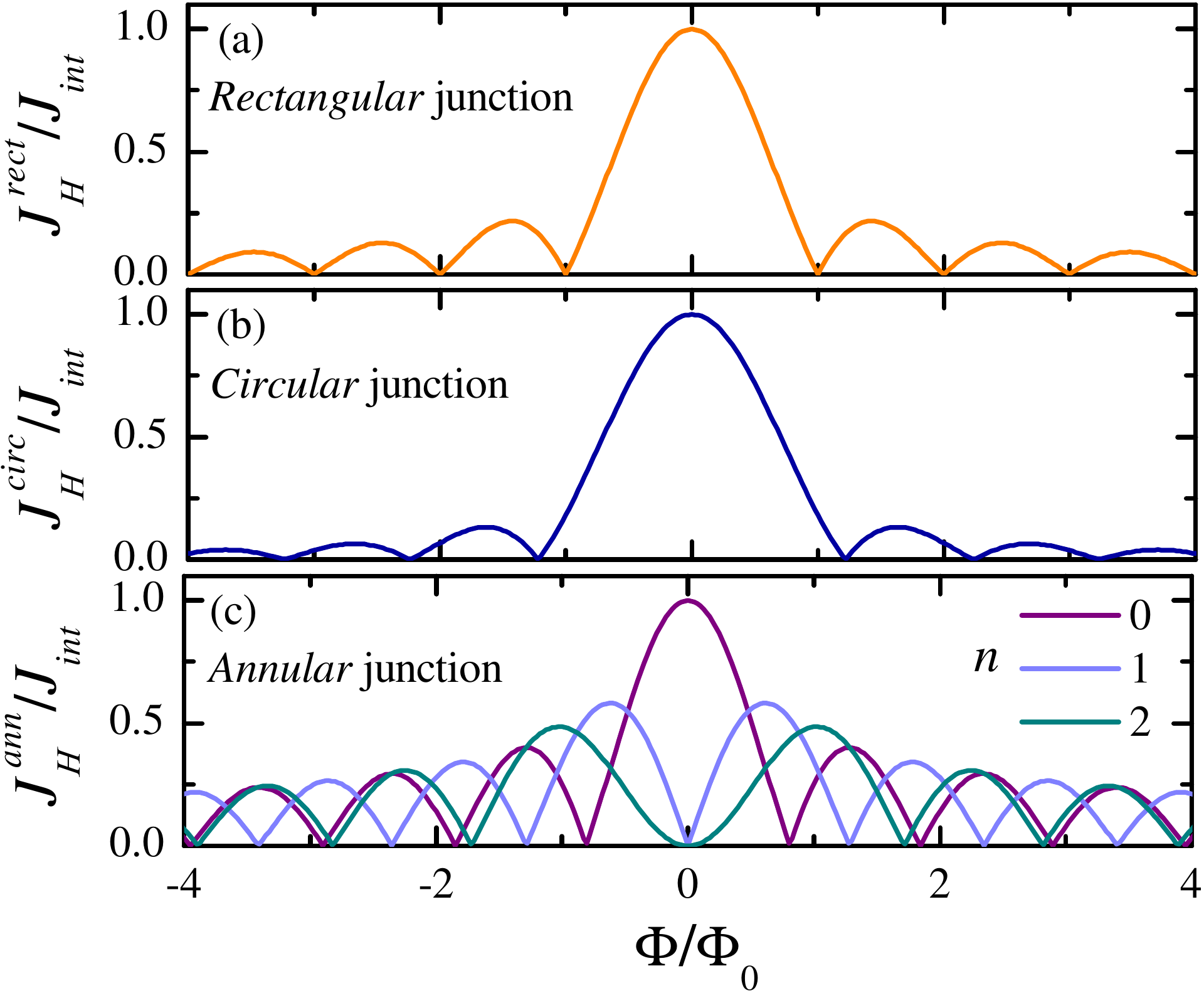} 
\caption{(Color online) Normalized phase-dependent  component of the heat current $J_{H}$ versus magnetic flux $\Phi$ calculated for a rectangular [(a)], circular [(b)], and annular [(c)] Josephson tunnel junction. In the curves of panel (c) we set $\alpha=0.9$, and $n$ indicates the number of fluxons trapped in the junction barrier.
}
\label{fig2}
\end{figure}
\section{Heat current in Josephson junctions with different geometries}
\label{differentgeo}
With the help of Eq. (\ref{heatcurrfin}) we can now determine the behavior of $J_{H}(T_1,T_2,H)$ for the three prototypical junction geometries sketched in Fig. (\ref{fig1}). 
In particular, we shall consider two well-known examples such as the \emph{rectangular} [see Fig. \ref{fig1}(b)] and \emph{circular} [see Fig. \ref{fig1}(c)] junction, and the more exotic \emph{annular} one [see Fig. \ref{fig1}(d)]. Annular junctions offer the possibility to investigate fluxons dynamics due to the absence of collisions with boundaries; yet, they provide fluxoid quantization thanks to their geometry which allows fluxons trapping. 
We assume that the total phase-dependent heat current is characterized by a uniform distribution, i.e., by a constant thermal current areal density $J_A(x,y,T_1,T_2)$ in Eq. (\ref{phasea}).  
$J_{H}$ can therefore be calculated for the three considered geometries by following, for instance, Refs. \cite{Tinkham,Barone}. 
In particular, for the rectangular junction,  the absolute value of the sine cardinal function is obtained,
\begin{equation}
J_{H}^{rect}(T_1,T_2,\Phi)=J_{int}(T_1,T_2)\left|\frac{\textrm{sin}(\pi\Phi/\Phi_0)}{(\pi\Phi/\Phi_0)}\right|, 
\end{equation}
where $J_{int}(T_1,T_2)=WLJ_A(T_1,T_2)$,
$\Phi=\mu_0 HLt_H$, $L$ is the junction length and $W$ its width.
For the circular geometry one gets the Airy diffraction pattern, 
\begin{equation}
J_{H}^{circ}(T_1,T_2,\Phi)=J_{int}(T_1,T_2)\left|\frac{J_1(\pi\Phi/\Phi_0)}{(\pi\Phi/2\Phi_0)}\right|,
\end{equation} 
where $J_{int}(T_1,T_2)=\pi R^2 J_A(T_1,T_2)$, $J_1(y)$ is the Bessel function of the first kind, $\Phi=2\mu_0 HRt_H$, and $R$ is the junction radius.
Finally, for the annular junction \cite{Martucciello1996,Nappi1997}, the phase-dependent component of the heat current takes the form
\begin{equation}
J_{H}^{ann}(T_1,T_2,\Phi)=\frac{2J_{int}(T_1,T_2)}{1-\alpha^2}\left|\int^1_\alpha dxxJ_n(x\pi \Phi/\Phi_0)\right|, 
\end{equation}
where $J_{int}(T_1,T_2)=\pi(R^2-r^2)J_A(T_1,T_2)$, $\Phi=2\mu_0 HRt_H$,
$\alpha=r/R$, $J_n(y)$ is the $n$th Bessel function of integer order, $R$ ($r$) is the external (internal)  radius, and $n=0,1,2,...$ is the number of $n$ trapped fluxons in the junction barrier.

Figure \ref{fig2} illustrates the behavior of $J_{H}$ for the three geometries. In particular, the curve displayed in Fig. \ref{fig2}(a) for the rectangular case shows the well-known Fraunhofer diffraction pattern analogous to that produced by light diffraction through a rectangular slit. In such a case, the heat current $J_{H}$ vanishes when the applied magnetic flux through the junction equals integer multiples of $\Phi_0$. Furthermore, the heat current is  rapidly damped by increasing the magnetic field falling asymptotically as $\Phi^{-1}$.\cite{Tinkham}
The behavior for a circular junction is displayed in Fig. \ref{fig2}(b). 
Here, the flux values where $J_{H}$ vanishes do not coincide anymore with multiples of $\Phi_0$, and $J_H$ falls more rapidly than in the rectangular junction case, i.e., as $\Phi^{-3/2}$.\cite{Tinkham}
Figure \ref{fig2}(c) shows $J_H$ for an annular junction. In particular, the heat current diffraction pattern is strongly $n$-dependent and, differently from the rectangular and circular case, $J_H$ decays in general more slowly. It is apparent that annular junctions may provide, in principle, enhanced flexibility to tailor the  heat current response. 
 
\begin{figure}[tb]
\includegraphics[width=\columnwidth]{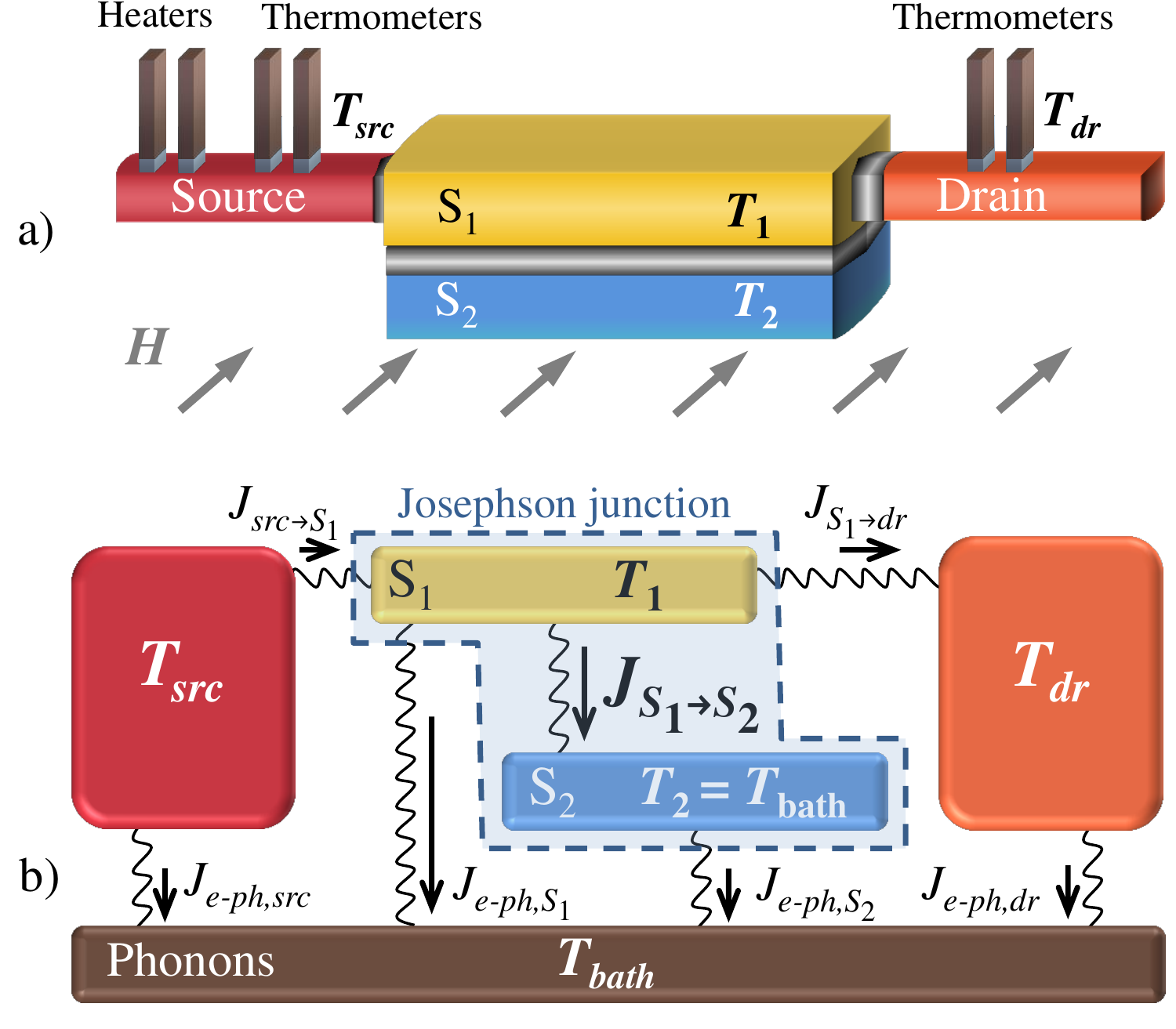} 
\caption{(Color online) (a) Possible experimental setup to demonstrate heat diffraction in a temperature-biased rectangular  Josephson junction. Source and drain normal-metal electrodes are tunnel-coupled to one of the junction electrodes (S$_1$). Superconducting tunnel junctions operated as heaters and thermometers are connected to source and drain. A static in-plane magnetic field $H$ is applied perpendicular to the S$_1$IS$_2$ junction.
(b) Thermal model describing the main heat exchange mechanisms existing in the structure shown in (a).}
\label{fig3}
\end{figure}
\section{Proposed experimental setup}
\label{experiment}
Demonstration of diffraction of thermal currents could be achieved in the setup shown in Fig. \ref{fig3}(a). It consists of two normal-metal source and drain electrodes tunnel-coupled via resistances $R_t$ to one electrode (S$_1$) of a Josephson junction which, for the sake of clarity, is assumed to be \emph{rectangular}. An in-plane static magnetic field $H$ is applied perpendicular to the Josephson weak-link.
Furthermore, superconducting probes tunnel-coupled to both source and drain electrodes either implement  heaters or allow accurate measurement of the electronic temperature in the leads \cite{Giazotto2006}. 
By intentionally heating electrons in the source up to $T_{src}$ yields a quasiparticle temperature $T_1>T_2$ in S$_1$,  therefore leading to a finite heat current $J_{S_1\rightarrow S_2}$. 
Yet, the latter can be modulated by the applied magnetic field. 
Measurement of the drain electron temperature ($T_{dr}$) would thus allow to assess heat diffraction. 

Drain temperature can be predicted by solving a couple of thermal balance equations accounting  for the main heat exchange mechanisms existing in the structure, according to the model shown in Fig. \ref{fig3}(b).
In particular, S$_1$ exchanges heat with source electrons at power $J_{src\rightarrow S_1}$, with drain at power $J_{S_1\rightarrow drain}$, and with quasiparticles in S$_2$ at power $J_{S_1 \rightarrow S_2}$. 
Furthermore, electrons in the structure exchange heat with lattice phonons residing at bath temperature $T_{bath}$, in particular, at power $J_{e-ph, S_1}$ in S$_1$, and at power $J_{e-ph,src}$  and $J_{e-ph,dr}$ in source and drain electrodes, respectively. 
Finally, we assume S$_2$ to be large enough to provide substantial electron-phonon coupling $J_{e-ph, S_2}$ so that its quasiparticles will reside at $T_{bath}$.
The electronic temperatures $T_1$ and $T_{dr}$ can therefore be determined under given conditions by solving the following system of thermal balance equations 
\begin{eqnarray}
-J_{src\rightarrow S_1}+J_{S_1 \rightarrow S_2}+J_{S_1\rightarrow drain}+J_{e-ph, S_1}&=&0\\ 
-J_{S_1\rightarrow drain}+J_{e-ph,dr}&=&0\nonumber
\end{eqnarray}
for S$_1$ and drain, respectively. 
In the above expressions,
$J_{S_1 \rightarrow S_2}(T_1,T_{bath},\Phi)=J_{qp}(T_1,T_{bath})-J_{H}^{rect}(T_1,T_{bath},\Phi)$,
$J_{src\rightarrow S_1}(T_{src},T_1)=\frac{1}{e^2 R_t} \int^{\infty}_{0} d\varepsilon \varepsilon \mathcal{N}_1 (\varepsilon,T_1)[f(\varepsilon,T_1)-f(\varepsilon,T_{src})]$,
$J_{S_1\rightarrow drain}(T_1,T_{dr})=\frac{1}{e^2 R_t} \int^{\infty}_{0} d\varepsilon \varepsilon \mathcal{N}_1 (\varepsilon,T_1)[f(\varepsilon,T_{dr})-f(\varepsilon,T_{1})]$, 
and $J_{e-ph,dr}=\Sigma_{dr}\mathcal{V}_{dr}(T^5_{dr}-T^5_{bath})$ \cite{Giazotto2006}, $\Sigma_{dr}$ and $\mathcal{V}_{dr}$ being the electron-phonon coupling constant and the volume of drain, respectively. Furthermore \cite{Timofeev2009}, 
\begin{eqnarray}
J_{e-ph,S_1}&=&-\frac{\Sigma_{S_1} \mathcal{V}_{S_1}}{96\zeta (5)k_B^5}\int^{\infty}_{-\infty}dEE\int^{\infty}_{-\infty}d\varepsilon \varepsilon^2\text{sign}(\varepsilon)M_{E,E+\varepsilon}\nonumber\\
&\times& [\text{coth}(\frac{\varepsilon}{2k_B T_{bath}})(f_E-f_{E+\varepsilon})-f_Ef_{E+\varepsilon}+1],
\label{eph}
\end{eqnarray}
where $f_E(T_1)=\text{tanh}(E/2k_B T_1)$, $M_{E,E'}(T_1)=\mathcal{N}_1(E,T_1)\mathcal{N}_1(E',T_1)[1-\Delta_1^2(T_1)/EE']$, $\Sigma_{S_1}$ is the electron-phonon coupling constant, and $\mathcal{V}_{S_1}$ is the volume of S$_1$. 
As a set of parameters representative for a realistic microstructure we choose $R_t=2\,\text{k}\Omega$, $R_J=500\,\Omega$, $\mathcal{V}_{dr}=10^{-20}$ m$^{-3}$, $\Sigma_{dr}=3\times 10^9$ WK$^{-5}$m$^{-3}$ (typical of Cu) \cite{Giazotto2006}, $\mathcal{V}_{S_1}=10^{-18}$ m$^{-3}$, $\Sigma_{S_1}=3\times 10^8$ WK$^{-5}$m$^{-3}$ and $\Delta_1(0)=\Delta_2(0)=200\,\mu$eV, the last two parameters typical of aluminum (Al) \cite{Giazotto2006}. Finally, our thermal model neglects both heat exchange with photons, owing to poor matching impedance \cite{schmidt,Meschke2006}, and pure phononic heat conduction \cite{Maki1965,giazottoexp2012}. 
\begin{figure}[tb]
\includegraphics[width=\columnwidth]{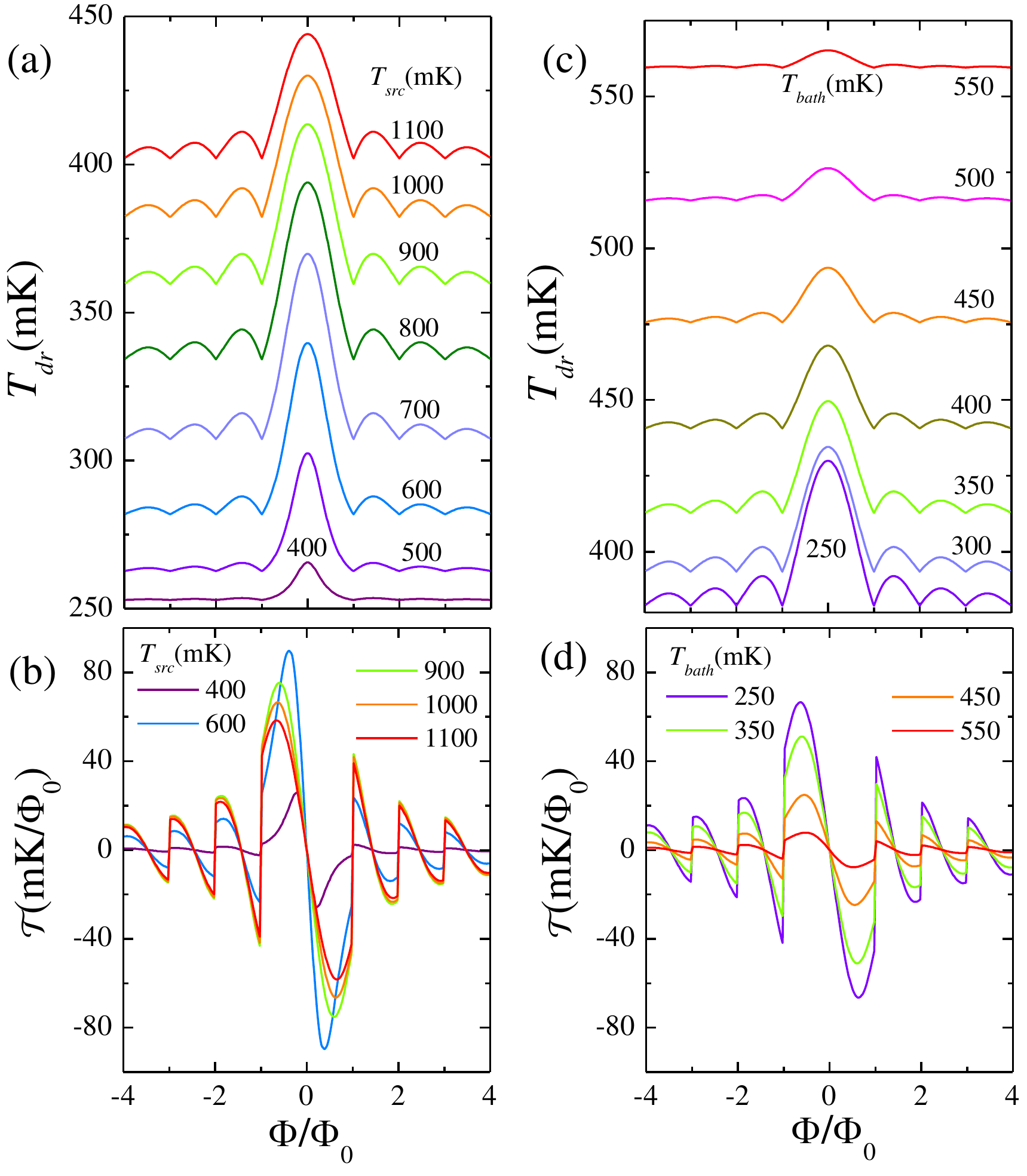} 
\caption{(Color online) (a) Drain temperature $T_{dr}$ vs $\Phi$ calculated at $T_{bath}=250$ mK for several values of source temperature $T_{src}$ for a structure based on a \emph{rectangular} Josephson junction. (b) Flux-to temperature transfer function $\mathcal{T}$ vs $\Phi$ calculated at 250 mK for a few selected values of $T_{src}$.
(c) $T_{dr}$ vs $\Phi$ calculated for a few values of $T_{bath}$ at $T_{src}=1$ K. (d) $\mathcal{T}$ vs $\Phi$ at a few selected $T_{bath}$ calculated for $T_{src}=1$ K.}
\label{fig4}
\end{figure}

The results of thermal balance equations for drain temperature are shown in Fig. \ref{fig4} \cite{Gamma}. In particular, panel (a) displays $T_{dr}$ vs $\Phi$ for different values of $T_{src}$ at $T_{bath}=250$ mK. 
As expected, $T_{dr}$ shows a response to magnetic flux resembling a Fraunhofer-like diffraction pattern. The minima appearing at integer values of $\Phi_0$ are the inequivocal manifestation of the above-described phase-slips. 
Increasing $T_{src}$ leads to a monotonic enhancement of the maximum of $T_{dr}$  at $\Phi=0$ which stems from an increased heat current flowing into drain electrode. 
Furthermore, the amplitude of $T_{dr}$ lobes follows a non-monotonic beahavior, initially increasing with source temperature, being maximized at intermediate temperatures, and finally decreasing at higher $T_{src}$ values. With the above-given structure parameters one would obtain a maximum peak-to-valley amplitude exceeding $\sim 60$ mK at $T_{src}\sim 700$ mK. 
By defining a figure of merit in the form of flux-to-temperature transfer coefficient, $\mathcal{T}=\partial T_{dr}/\partial \Phi$, we get that $\mathcal{T}$ as large as $\sim 90$ mK$/\Phi_0$ could be achieved at $T_{src}=600$ mK in the present structure [see Fig. \ref{fig4}(b)]. Moreover, the transfer coefficient clearly demonstrates the non-monotonicity of the amplitude of drain temperature lobes as a function of $T_{src}$.

The impact of bath temperature on the structure response is shown in Fig. \ref{fig4}(c) where $T_{dr}$ is plotted against $\Phi$ for a few $T_{bath}$ values at fixed $T_{src}=1$ K. In particular, by increasing $T_{bath}$ leads to a smearing of drain temperature joined with a reduction of the lobes amplitude. This originates from
both reduced temperature drop across the Josephson junction and enhanced electron-phonon relaxation in S$_1$ and drain at higher $T_{bath}$. 
We notice that already at 550 mK the temperature diffraction pattern is somewhat suppressed for a structure realized according to the chosen parameters. The drain temperature behavior as a function of $T_{bath}$ directly reflects on the transfer coefficient $\mathcal{T}(\Phi)$ [see Fig. \ref{fig4}(d)] which is calculated for a few selected values of $T_{bath}$.  
 
We finally notice that the temperature diffraction patterns shown in Figs. \ref{fig4} implicitly assume the presence of the $\pi$ slips and, therefore, the same heat diffraction measure can be considered as a proof of the existence of such phase slips.

\section{Summary}
\label{summary}
In summary, we have investigated thermal transport in temperature-biased extended Josephson tunnel junctions under the influence of an  in-plane magnetic field. 
We have shown, in particular, that the heat current through the junction displays \emph{coherent diffraction}, in full analogy with the Josephson critical current. In an electrically-open junction configuration,
minimization of the Josephson coupling energy imposes the quantum phase difference across the junction to undergo $\pi$ slips in suitable magnetic flux intervals, the latter depending on the specific junction geometry. 
Finally, we have proposed and analyzed a hybrid superconducting microstructure, easily implementable with current technology, which would allow to demonstrate diffraction of thermal currents. 
We wish further  to stress that  the described temperature detection is uniquely suited to reveal the hidden physical properties of the quantum phase in electrically-open tunnel junctions of whatever geometry otherwise more difficult to access with electric-type transport measurement.
The effects here predicted could serve to enhance the flexibility to master thermal currents in emerging coherent caloritronic  nanocircuitry.

\acknowledgments
We would like to thanks C. Altimiras for useful discussions.
F.G. and M.J.M.-P. acknowledges the FP7 program No. 228464 ``MICROKELVIN'', the Italian Ministry of Defense through the PNRM project ``TERASUPER'', and  the Marie Curie Initial Training Action (ITN) Q-NET 264034 for partial financial support. 
P.S. acknowledges financial support from FIRB - Futuro in Ricerca 2012 under Grant No. RBFR1236VV HybridNanoDev.


\begin{thebibliography}{99}
\bibitem{Giazotto2006} 
	F. Giazotto, T. T. Heikkil\"a, A. Luukanen, A. M. Savin, and J. P. Pekola, Rev. Mod. Phys. \textbf{78}, 217 (2006).
\bibitem{Dubi2011} 
	Y. Dubi and M. Di Ventra, Rev. Mod. Phys. \textbf{83}, 131 (2011). 
\bibitem{heattransistor} 
	O.-P. Saira, M. Meschke, F. Giazotto, A. M. Savin, M. M\"ott\"onen, and J. P. Pekola, Phys. Rev. Lett. \textbf{99}, 027203 (2007).
\bibitem{ser} 
	J. P. Pekola, F. Giazotto, and O.-P. Saira, Phys. Rev. Lett. \textbf{98}, 037201 (2007).
\bibitem{Meschke2006}
	M. Meschke, W. Guichard, and J. P. Pekola, Nature \textbf{444}, 187 (2006).
\bibitem{Vinokur2003}
	E. V. Bezuglyi and V. Vinokur, Phys. Rev. Lett. \textbf{91}, 137002 (2003).
\bibitem{Eom1998}
	J. Eom, C.-J. Chien, and V. Chandrasekhar, Phys. Rev. Lett. \textbf{81}, 437 (1998).
\bibitem{Chandrasekhar2009}
	V. Chandrasekhar, Supercond. Sci. Technol. \textbf{22}, 083001 (2009).
\bibitem{Ryazanov1982}
	V. V. Ryazanov and V. V. Schmidt, Solid State Commun. \textbf{42}, 733 (1982).
\bibitem{Panaitov1984}
	G. I. Panaitov, V. V. Ryazanov, and V. V. Schmidt, Phys. Lett. \textbf{100}, 301 (1984).
\bibitem{virtanen2007}
	P. Virtanen and T. T. Heikkil\"a, Appl. Phys. A \textbf{89}, 625 (2007). 
\bibitem{Martinez2013}
	M. J. Mart\'inez-P\'erez and F. Giazotto, Appl. Phys. Lett. \textbf{102}, 182602 (2013).
\bibitem{Giazotto2008}
	F. Giazotto, T. T. Heikkil\"a, G. P. Pepe, P. Helisto, A. Luukanen, and J. P. Pekola, Appl. Phys. Lett. \textbf{92}, 162507 (2008).
\bibitem{Giazotto2002}
	F. Giazotto, F. Taddei, R. Fazio, and F. Beltram, Appl. Phys. Lett. \textbf{80}, 3784 (2002).
\bibitem{Maki1965} 
	K. Maki and A. Griffin, Phys. Rev. Lett. {\bf 15}, 921 (1965).
\bibitem{Guttman97} 
	G. D. Guttman, B. Nathanson, E. Ben-Jacob, and D. J. Bergman, Phys. Rev. B \textbf{55}, 3849 (1997).
\bibitem{Guttman98} 
	G. D. Guttman, E. Ben-Jacob, and D. J. Bergman, Phys. Rev. B \textbf{57}, 2717 (1998).
\bibitem{Zhao2003} 
	E. Zhao, T. L\"ofwander, and J. A. Sauls, Phys. Rev. Lett. \textbf{91}, 077003 (2003).
\bibitem{Zhao2004} 
	E. Zhao, T. L\"ofwander, and J. A. Sauls, Phys. Rev. B \textbf{69}, 134503 (2004).
\bibitem{Golubev2013}
	D. Golubev, T. Faivre, and J. P. Pekola, Phys. Rev. B \textbf{87}, 094522 (2013).
\bibitem{giazotto2012} 
	F. Giazotto and M. J. Mart\'inez-P\'erez, Appl. Phys. Lett. \textbf{101}, 102601 (2012).
\bibitem{martinez2012} 
	M. J. Mart\'inez-P\'erez and F. Giazotto, Appl. Phys. Lett. \textbf{102}, 092602 (2013).
\bibitem{giazottoexp2012} 
	F. Giazotto and M. J. Mart\'inez-P\'erez, Nature \textbf{492}, 401 (2012).
\bibitem{simmonds2012}
	R. W. Simmonds, Nature \textbf{492}, 358 (2012).
\bibitem{Langer1967} J.S. Langer, V. Ambegaokar, Phys. Rev. {\bf 164}, 498 (1967).
\bibitem{Zaikin1997} A.D. Zaikin, D.S. Golubev, A. van Otterlo, G.T. Zimanyi, Phys. Rev. Lett. {\bf 78}, 1552 (1997).
\bibitem{astafiev2012}
O. V. Astafiev,	L. B. Ioffe, S. Kafanov, Yu. A. Pashkin,	K. Yu. Arutyunov,	D. Shahar,	O. Cohen, and J. S. Tsai,     Nature \textbf{484}, 355 (2012).
\bibitem{arutyunov08}  
K.Yu. Arutyunov, D.S. Golubev and A.D. Zaikin, Physics Reports \textbf{464}, 1 (2008).
\bibitem{magneticlength}
When this condition is no longer satisfied the total magnetic penetration depth has to be replaced with an effective thickness ($\tilde{t}_H$) given by $\tilde{t}_H=\lambda_1\textrm{tanh}(t_1/2\lambda_1)+\lambda_2\textrm{tanh}(t_2/2\lambda_2)+d$ \cite{Weihnacht1969}.   
\bibitem{Weihnacht1969}
  M. Weihnacht, Phys. Status Solidi \textbf{32}, K169 (1969).
\bibitem{Frank1997}
	B. Frank and W. Krech, Phys. Lett. A \textbf{235}, 281 (1997).
\bibitem{Tinkham}
	M. Tinkham, \emph{Introduction to Superconductivity 2nd Edn.} (McGraw-Hill, New York, 1996).
\bibitem{Barone}
	A. Barone and G. Patern\'o, \emph{Physics and Applications of the Josephson Effect} (Wiley, New York, 1982).
\bibitem{Martucciello1996}
	N. Martucciello and R. Monaco, Phys. Rev. B \textbf{53}, 3471 (1996).
\bibitem{kuplevakhsky06} 
S. V. Kuplevakhsky and A. M. Glukhov, Phys. Rev. B \textbf{ 73}, 024513 (2006).
\bibitem{Nappi1997}
	C. Nappi, Phys. Rev. B \textbf{55}, 82 (1997).
\bibitem{Gamma}
	Throughout our analysis we added a small imaginary part to the energy in $\mathcal{N}_i(\varepsilon)$ and $\mathcal{M}_i(\varepsilon)$ to account for smearing, i.e., $\varepsilon \rightarrow \varepsilon+i\gamma$, where $\gamma=10^{-5}\Delta_1(0)$ \cite{Martinez2013,Dynes1984,Pekola2004,Pekola2010}.
	\bibitem{Dynes1984}
	R. C. Dynes, J. P. Garno, G. B. Hertel, and T. P. Orlando, Phys. Rev. Lett. \textbf{53}, 2437 (1984).
\bibitem{Pekola2004}
	J. P. Pekola, T. T. Heikkil\"a, A. M. Savin, J. T. Flyktman, F. Giazotto, and F. W. J. Hekking, Phys. Rev. Lett. \textbf{92}, 056804 (2004).
\bibitem{Pekola2010}
	J. P. Pekola, V. F. Maisi, S. Kafanov, N. Chekurov, A. Kemppinen, Yu. A. Pashkin, O.-P. Saira, M. M\"ott\"onen, and J. S. Tsai, Phys. Rev. Lett. \textbf{105}, 026803 (2010).
\bibitem{Timofeev2009}
	A. V. Timofeev, C. P. Garcia, N. B. Kopnin, A. M. Savin, M. Meschke, F. Giazotto, and J. P. Pekola, Phys. Rev. Lett. \textbf{102}, 017003 (2009).
\bibitem{schmidt}
	D. R. Schmidt, R. J. Schoelkopf, and A. N. Cleland, Phys. Rev. Lett. \textbf{93}, 045901 (2004).




\end{thebibliography}
\end{document}